\documentclass[article]{elsarticle}
\usepackage{a4wide}
\usepackage{natbib} 			
\usepackage{graphicx}
\usepackage{amsfonts}
\usepackage{amsmath}
\usepackage{booktabs}
\usepackage{epsfig}
\usepackage[colorlinks]{hyperref}
\hypersetup{citecolor=black,linkcolor= black}

\begin{document}
\title{Mixed-correlated ARFIMA processes for power-law cross-correlations}
\author{Ladislav Kristoufek}
\ead{kristouf@utia.cas.cz}
\address{Institute of Information Theory and Automation, Academy of Sciences of the Czech Republic, Pod Vodarenskou Vezi 4, 182 08, Prague 8, Czech Republic\\
Institute of Economic Studies, Faculty of Social Sciences, Charles University, Opletalova 26, 110 00, Prague 1, Czech Republic
}

\begin{abstract}
We introduce a general framework of the Mixed-correlated ARFIMA (MC-ARFIMA) processes which allows for various specifications of univariate and bivariate long-term memory. Apart from a standard case when $H_{xy}=\frac{1}{2}(H_x+H_y)$, MC-ARFIMA also allows for processes with $H_{xy}<\frac{1}{2}(H_x+H_y)$ but also for long-range correlated processes which are either short-range cross-correlated or simply correlated. The major contribution of MC-ARFIMA lays in the fact that the processes have well-defined asymptotic properties for $H_x$, $H_y$ and $H_{xy}$, which are derived in the paper, so that the processes can be used in simulation studies comparing various estimators of the bivariate Hurst exponent $H_{xy}$. Moreover, the framework allows for modeling of processes which are found to have $H_{xy}<\frac{1}{2}(H_x+H_y)$.
\end{abstract}

\begin{keyword}
power-law cross-correlations, long-term memory, econophysics
\end{keyword}

\journal{Physica A}

\maketitle

\textit{PACS codes: 05.45.-a, 05.45.Tp, 89.65.Gh}\\

\section{Introduction}

Studying long-range correlations has become a stable part of econophysics in recent years. Long-term memory has been studied for stock indices \citep{DiMatteo2003,DiMatteo2005,Barunik2010,Kristoufek2010a,Kristoufek2010b}, bonds \citep{Carbone2004,DiMatteo2005}, exchange rates \citep{Vandewalle1998,Norouzzadeh2006}, commodities \citep{Alvarez-Ramirez2002,Alvarez-Ramirez2009,Power2010}, interest rates \citep{Cajueiro2007,DiMatteo2007} and others. Most recently, a focus has been put on analysis of long-range (power-law) cross-correlations as an addition to more traditional research of power-law auto-correlations. Similarly to the power-law auto-correlation case, we assume that a pair of time series is long-range cross-correlated if their cross-correlation function $\rho_{xy}(k)$ with lag $k$ follows an asymptotic power law so that $\rho_{xy}(k) \propto k^{2H_{xy}-2}$ as $k\rightarrow +\infty$ where $H_{xy}$ is the bivariate Hurst exponent. For $H_{xy}>0.5$, we have cross-persistent processes which possess non-zero cross-correlations even for very long lags. Contrary to the univariate case, we can have both positively and negatively cross-persistent process with $H_{xy}>0.5$ due to different properties of the auto- and cross-correlation functions.

As the financial time series provide sufficient number of observations and they have been shown to possess long-range auto-correlations (volatility, traded volume, signs of changes and absolute returns to name the most frequently studied ones), these also create an appropriate setting for the power-law cross-correlations analysis \cite{Podobnik2009,Podobnik2009a,SiqueiraJr.2010,He2011,He2011b,Ma2013,Wang2013,Wang2013a}. For estimation of the bivariate Hurst exponent $H_{xy}$, several estimators have been proposed. Podobnik \& Stanley \cite{Podobnik2008} propose the detrended cross-correlation analysis (DCCA) as a bivariate generalization of the detrended fluctuation analysis (DFA) \cite{Peng1994}. DCCA has since been expanded to various specifications \cite{Gu2010,Jiang2011}. Zhou \citep{Zhou2008} then generalizes the method to the multifractal setting with the multifractal detrended cross-correlations analysis (MF-DXA). Kristoufek \cite{Kristoufek2011} generalizes the height-height correlation analysis \cite{Barabasi1991a,Barabasi1991b} and the generalized Hurst exponent approach \cite{DiMatteo2003} into the multifractal height cross-correlation analysis (MF-HXA). The detrending moving average (DMA) \cite{Alessio2002} is generalized by He \& Chen \cite{He2011a} forming the detrended moving-average cross-correlation analysis (DMCA). The most recent estimator from the time domain is proposed by Wang \textit{et al.} \cite{Wang2012} -- multifractal cross-correlation analysis based on statistical moments (MFSMXA). However, relatively little attention has been given to processes which can be characterized as cross-persistent with a specific bivariate Hurst exponent.

Several processes that possess such long-term correlations have been proposed in the literature. The most frequently discussed and applied ones are the multivariate generalizations of the well-established fractionally integrated ARMA processes (usually labeled as FARIMA and ARFIMA) -- VARFIMA or MVARFIMA processes \cite{Shimotsu2007,Tsay2010,Nielsen2011} -- and fractional Gaussian noise processes or fractional Brownian motions, which are their integrated version (these are labeled as fGn and fBm in the literature, respectively). The construction of the multivariate ARFIMA process implies that the bivariate Hurst exponent is the average of the separate Hurst exponents \citep{Nielsen2011}. The same property holds for the fractional Brownian motion \citep{Amblard2011}. The long-range cross-correlations thus simply arise from the specification of these processes.

Lobato \cite{Lobato1997} and then in some detail Sela \& Hurvich \cite{Sela2009} discuss two types of fractionally integrated models -- VARFI and FIVAR. VARFI is a vector autoregressive model with fractionally integrated innovations (or error terms). The pairwise processes are then long-range correlated but only short-range cross-correlated so that the bivariate Hurst exponent is equal to 0.5 and is thus lower than the average of the separate Hurst exponents. Reversely, FIVAR consists of fractionally integrated processes with innovations that come from a VAR model. The pairwise processes are then both long-range correlated and long-range cross-correlated with the bivariate Hurst exponent being equal to the average of the separate Hurst exponents as for the previous cases. 

Nielsen \cite{Nielsen2004a} and Sela \cite{Sela2010} discuss the case of the fractional cointegration, i.e. the case when both processes are fractionally integrated of the same order and there exists a linear combination of them which is stationary and fractionally integrated of a lower order, in the bivariate long-term memory setting. Both authors show that the coherence of the processes is equal to unity which implies that the bivariate Hurst exponent is the same as the separate Hurst exponents and so again is equal to their average.

All of the above mentioned processes possibly yield cross-persistence but only with the bivariate Hurst exponent equal to the average of the Hurst exponents of the separate processes. There are only two studies which propose models with $H_{xy}$ different from $\frac{1}{2}(H_x+H_y)$. Sela \& Hurvich \cite{Sela2012} propose an anti-cointegration model, which is in fact a linear combination of ARFIMA processes with a subset of innovations (but not all pairs) being identical across the two processes. The model allows to control the separate Hurst exponents as well as the bivariate Hurst exponent as long as it is lower or equal to the average of the separate parameters. Podobnik \textit{et al.} \cite{Podobnik2008a} introduce two-component ARFIMA processes, which are based on ARFIMA-like mixing of two processes. Unfortunately, the authors neither provide any clue how to control the bivariate parameter $H_{xy}$ nor is it evident whether the processes are even stationary.

In this paper, we introduce a new kind of a bivariate process which we call the mixed-correlated ARFIMA process. The process allows to control for the separate and bivariate Hurst exponents as long as the bivariate one is not higher than the average of the separate ones, and additionally allows for short-range dependence as well.

\section{Mixed-correlated ARFIMA framework}

We start with a general framework of a bivariate series where each of the series consists of a linear combination of two ARFIMA(0,$d$,0) processes so that
\begin{gather}
x_t=\alpha\sum_{n=0}^{+\infty}{a_n(d_1)\varepsilon_{1,t-n}}+\beta\sum_{n=0}^{+\infty}{a_n(d_2)\varepsilon_{2,t-n}} \nonumber \\
y_t=\gamma\sum_{n=0}^{+\infty}{a_n(d_3)\varepsilon_{3,t-n}}+\delta\sum_{n=0}^{+\infty}{a_n(d_4)\varepsilon_{4,t-n}}
\label{eq:ARFIMA_LC}
\end{gather}
where 
\begin{equation}
a_n(d)=\frac{\Gamma(n+d)}{\Gamma(n+1)\Gamma(d)}
\end{equation}
and innovations are characterized by
\begin{gather}
\langle \varepsilon_{i,t} \rangle = 0\text{ for }i=1,2,3,4 \nonumber\\
\langle \varepsilon_{i,t}^2 \rangle = \sigma_{\varepsilon_i}^2\text{ for }i=1,2,3,4 \nonumber\\
 \langle \varepsilon_{i,t}\varepsilon_{j,t-n} \rangle = 0\text{ for }n \ne 0\text{ and }i,j=1,2,3,4 \nonumber\\
\langle \varepsilon_{i,t}\varepsilon_{j,t} \rangle = \sigma_{ij}\text{ for }i,j=1,2,3,4\text{ and }i\ne j.
\end{gather}

In words, we have two processes and each one is a linear combination of two long-range correlated processes with possibly correlated innovations. Note that the separate long-term memory parameters $d_1,d_2,d_3,d_4$ can vary or be the same. We call the set of processes $\{x_t\}$ and $\{y_t\}$ as the mixed-correlated ARFIMA processes (MC-ARFIMA). As MC-ARFIMA is a new kind of process not discussed in the literature, even though these can be seen as a generalization of the anti-cointegration model of Sela \& Hurvich \cite{Sela2012}, we shortly discuss its stationarity. For the wide-sense stationarity, it suffices to state that both $\{x_t\}$ and $\{y_t\}$ are linear combinations of two ARFIMA(0,$d$,0) processes with correlated innovations which are wide-sense stationary so that MC-ARFIMA processes are stationary as long as $0\le d_1,d_2,d_3,d_4<0.5$ \citep{Samorodnitsky2006}. Evidently, we have $\langle x_t \rangle=\langle y_t \rangle=0$ and both processes have finite variance, i.e. $\langle x_t^2 \rangle \equiv \sigma_x^2<+\infty$ and $\langle y_t^2 \rangle \equiv \sigma_y^2<+\infty$ since the separate ARFIMA(0,$d$,0) processes have zero means and finite variances. As ARFIMA(0,$d$,0) processes are long-range correlated, their linear combination is also long-range correlated. The higher $d$ will dominate in the linear combination so that process $\{x_t\}$ is integrated of order $\max(d_1,d_2)$ and $\{y_t\}$ of order $\max(d_3,d_4)$. The separate processes are thus wide-sense stationary.

To show that $\{x_t\}$ and $\{y_t\}$ are also jointly wide-sense stationary, we need to show that $\rho_{xy}(k)$ does not depend on $t$. In Appendix, we show a pattern of behavior of the cross-correlation function and it is evident that the cross-correlation function is dependent only on the parameters $d_1$, $d_2$, $d_3$, $d_4$, $\alpha$, $\beta$, $\gamma$, $\delta$ and $\sigma_{ij}$ (with $i,j=1,2,3,4$) and the processes $\{x_t\}$ and $\{y_t\}$ are thus also jointly wide-sense stationary. Based on the cross-correlation structure, the cross-power spectrum can be written as
\footnotesize
\begin{multline}
f_{xy}(\lambda)=\frac{\alpha\gamma\sigma_{13}}{2\pi}\sum_{k=0}^{+\infty}\sum_{l=0}^{+\infty}{a_k(d_1)a_l(d_3)\exp(i(k-l)\lambda)}+\frac{\alpha\delta \sigma_{14}}{2\pi}\sum_{k=0}^{+\infty}\sum_{l=0}^{+\infty}{a_k(d_1)a_l(d_4)\exp(i(k-l)\lambda)}+\\
\frac{\beta\gamma \sigma_{23}}{2\pi}\sum_{k=0}^{+\infty}\sum_{l=0}^{+\infty}{a_k(d_2)a_l(d_3)\exp(i(k-l)\lambda)}+\frac{\beta\delta\sigma_{24}}{2\pi}\sum_{k=0}^{+\infty}\sum_{l=0}^{+\infty}{a_k(d_2)a_l(d_4)\exp(i(k-l)\lambda)}=\\
\frac{1}{2\pi}\Big[\alpha\gamma\sigma_{13}\left(1-\exp(i\lambda)\right)^{-d_1}\left(1-\exp(-i\lambda)\right)^{-d_3}+\alpha\delta\sigma_{14}\left(1-\exp(i\lambda)\right)^{-d_1}\left(1-\exp(-i\lambda)\right)^{-d_4}+\\
\beta\gamma\sigma_{23}\left(1-\exp(i\lambda)\right)^{-d_2}\left(1-\exp(-i\lambda)\right)^{-d_3}+\beta\delta\sigma_{24}\left(1-\exp(i\lambda)\right)^{-d_2}\left(1-\exp(-i\lambda)\right)^{-d_4}\Big].
\end{multline}

\normalsize
Using the inverse Fourier transform and the Dirac delta function, we get
\footnotesize
\begin{multline}
\rho_{xy}(n)=\frac{\alpha\gamma\sigma_{13}}{\sigma_x\sigma_y}\sum_{k=0}^{+\infty}\sum_{l=0}^{+\infty}{a_k(d_1)a_l(d_3)\delta(n+k-l)}+\frac{\alpha\delta\sigma_{14}}{\sigma_x\sigma_y}\sum_{k=0}^{+\infty}\sum_{l=0}^{+\infty}{a_k(d_1)a_l(d_4)\delta(n+k-l)}+\\
\frac{\beta\gamma\sigma_{23}}{\sigma_x\sigma_y}\sum_{k=0}^{+\infty}\sum_{l=0}^{+\infty}{a_k(d_2)a_l(d_3)\delta(n+k-l)}+\frac{\beta\delta\sigma_{24}}{\sigma_x\sigma_y}\sum_{k=0}^{+\infty}\sum_{l=0}^{+\infty}{a_k(d_2)a_l(d_4)\delta(n+k-l)}=\\
\frac{\alpha\gamma\sigma_{13}}{\sigma_x\sigma_y}\underbrace{\sum_{k=0}^{+\infty}{a_k(d_1)a_{n+k}(d_3)}}_{\approx \int_{0}^{+\infty}{k^{d_1-1}(n+k)^{d_3-1}dk}\propto n^{d_1+d_3-1}}+\frac{\alpha\delta\sigma_{14}}{\sigma_x\sigma_y}\underbrace{\sum_{k=0}^{+\infty}{a_k(d_1)a_{n+k}(d_4)}}_{\approx \int_{0}^{+\infty}{k^{d_1-1}(n+k)^{d_4-1}dk}\propto n^{d_1+d_4-1}}+\\
\frac{\beta\gamma\sigma_{23}}{\sigma_x\sigma_y}\underbrace{\sum_{k=0}^{+\infty}{a_k(d_2)a_{n+k}(d_3)}}_{\approx \int_{0}^{+\infty}{k^{d_2-1}(n+k)^{d_3-1}dk}\propto n^{d_2+d_3-1}}+\frac{\beta\delta\sigma_{24}}{\sigma_x\sigma_y}\underbrace{\sum_{k=0}^{+\infty}{a_k(d_2)a_{n+k}(d_4)}}_{\approx \int_{0}^{+\infty}{k^{d_2-1}(n+k)^{d_4-1}dk}\propto n^{d_2+d_4-1}}.
\label{eq:rhon_MC-ARFIMA}
\end{multline}
\normalsize
The results are obtained by using the Stirling's approximation and by approximating the infinite sum by the definite integrals. As we are interested in the asymptotic case $n\rightarrow +\infty$, the scaling of $\rho_{xy}(n)$ will be dominated by the highest exponent. This leads us to several interesting settings.

Firstly, let's have $\alpha,\beta,\gamma,\delta \ne 0$ and $\sigma_{ij}\ne 0$ for all $i,j=1,2,3,4$. Labeling $H_i=d_i+0.5$ for $i=1,2,3,4$, we have \begin{gather}
H_x=\max(H_1,H_2) \nonumber  \\ 
H_y=\max(H_3,H_4)
\end{gather}
\begin{multline}
H_{xy}=\frac{\max(H_1+H_2,H_1+H_4,H_2+H_3,H_2+H_4)}{2}=\\
\frac{\max(H_1,H_2)+\max(H_3,H_4)}{2}=\frac{H_x+H_y}{2}.
\end{multline}
Therefore, if the innovations are correlated without restriction, we arrive at $H_{xy}=\frac{1}{2}(H_x+H_y)$.

\begin{figure}[!htbp]
\begin{center}
\begin{tabular}{cc}
\includegraphics[width=40mm]{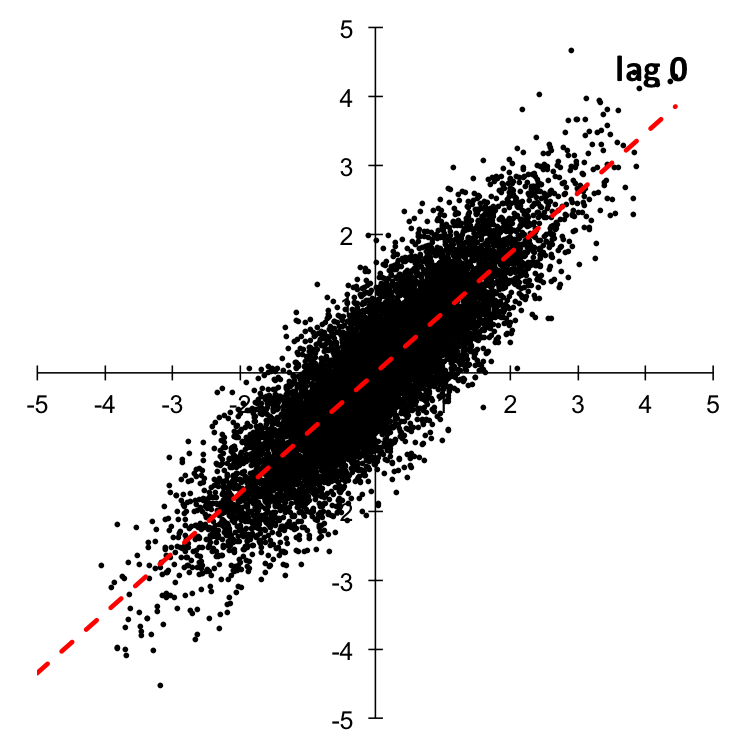}&\includegraphics[width=40mm]{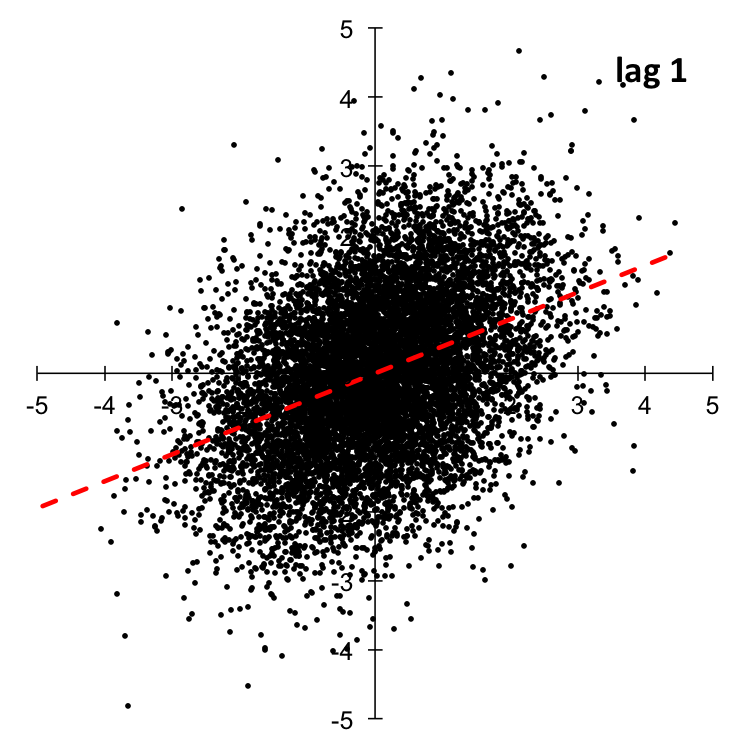}\\
\includegraphics[width=40mm]{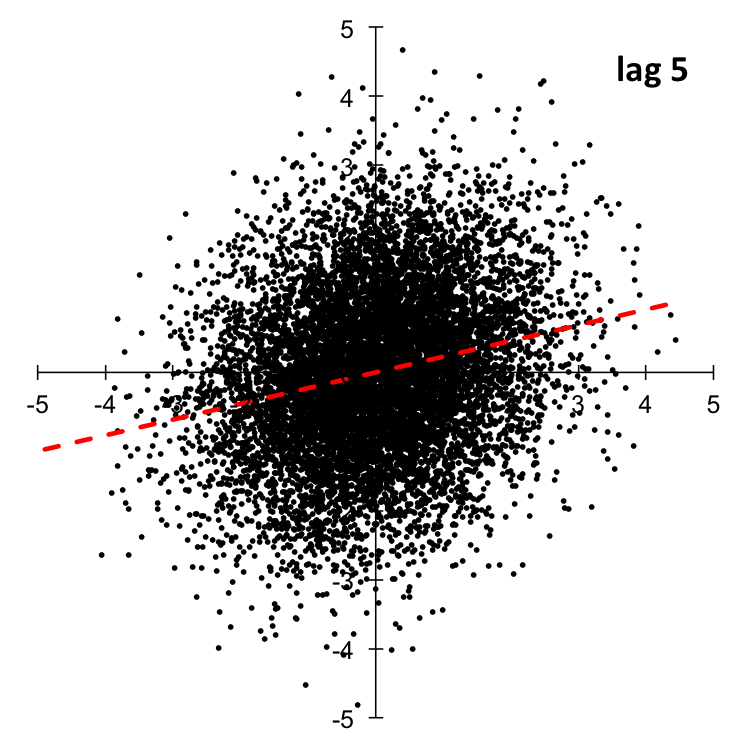}&\includegraphics[width=40mm]{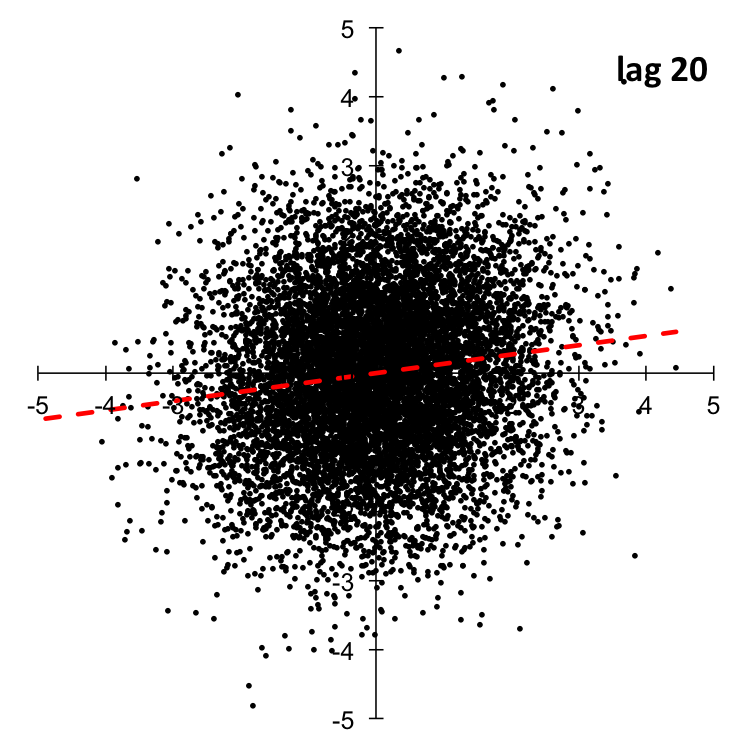}
\end{tabular}
\caption{\textbf{Cross-correlations decay of Model 1.} \footnotesize{Behavior of cross-correlations of selected lags illustrates a slow decay. Even for the twentieth lag, the relationship between lagged series is evident. Red dashed lines represent the least squares fits.}\label{fig:Mixed_ARFIMA1}}
\end{center}\vspace{-0.5cm}
\end{figure}

Secondly, let's again have $\alpha,\beta,\gamma,\delta \ne 0$ and without loss on generality, let's have $\max(H_1,H_2)=H_1$ and $\max(H_3,H_4)=H_4$ so that $\{x_t\}$ is integrated of order $d_1$ (with $H_x=0.5+d_1$) and $\{y_t\}$ of order $d_4$ (with $H_y=0.5+d_4$). Moreover, assume that $\sigma_{23}=\sigma_{32}\ne0$ and all the other covariances are equal to zero. From Eq. \ref{eq:rhon_MC-ARFIMA}, this implies 
\begin{equation}
H_{xy}=\frac{H_2+H_3}{2}\le \frac{H_x+H_y}{2}=\frac{H_1+H_4}{2}=\frac{\max(H_1,H_2)+\max(H_3,H_4)}{2}.
\end{equation}
The equality holds only if $H_1=H_2$ and $H_3=H_4$. For the other cases, it implies that the bivariate Hurst exponent $H_{xy}$ is not equal to the average of the univariate Hurst exponents $H_x$ and $H_y$ while still showing long-range cross-correlations, i.e. without $H_{xy}=0.5$. Processes $\{x_t\}$ and $\{y_t\}$ are thus long-range cross-correlated but possess the power-law coherency in a similar manner as the anti-cointegration model of Sela \& Hurvich \cite{Sela2012}.

\begin{figure}[!htbp]
\begin{center}
\begin{tabular}{c}
\includegraphics[width=80mm]{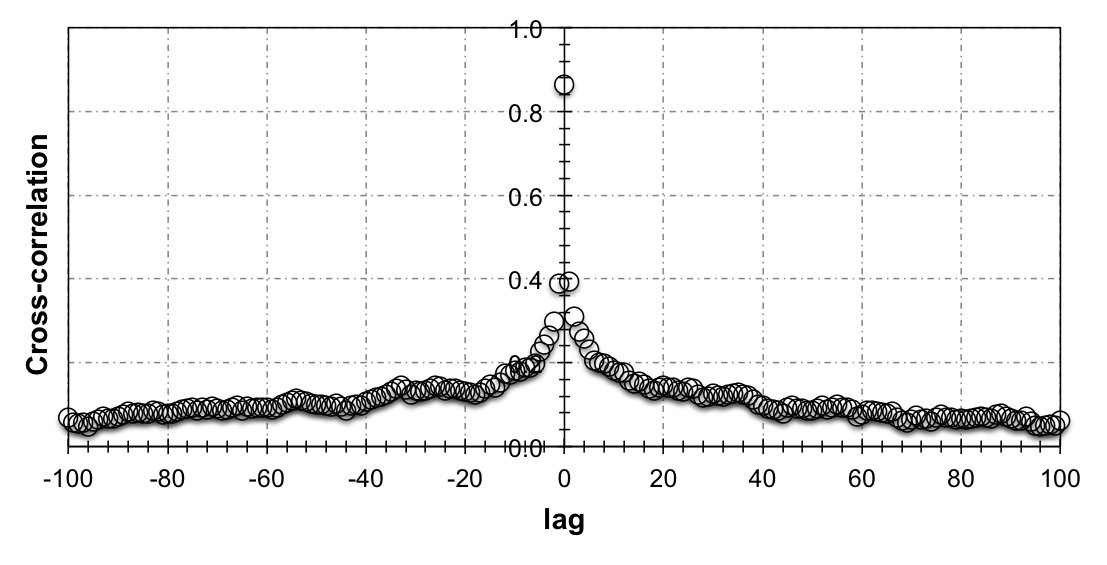}
\end{tabular}
\caption{\textbf{Cross-correlation function of Model 1.} \footnotesize{Cross-correlation function shows a very slow decay for both negative and positive lags.}\label{fig:Mixed_ARFIMA1_CCF}}
\end{center}\vspace{-0.5cm}
\end{figure}

Apart from the framework introduced above, we can slightly adjust the setting in the following way:
\begin{gather}
x_t=\alpha\sum_{n=0}^{+\infty}{a_n(d_1)\varepsilon_{1,t-n}}+\beta\varepsilon_{2,t} \nonumber \\
y_t=\gamma\varepsilon_{3,t}+\delta\sum_{n=0}^{+\infty}{a_n(d_4)\varepsilon_{4,t-n}}.
\label{eq:ARFIMA_Noise}
\end{gather}
We thus again have $\{x_t\}$ with the long-term memory parameter $d_1$ (with $H_x=0.5+d_1$) and $\{y_t\}$ with $d_4$ (with $H_y=0.5+d_4$) but $H_{xy}=0.5$ since it can be easily shown that $\rho_{xy}(0)=\frac{\sigma_{23}}{\sigma_x\sigma_y}$ and $\rho_{xy}(k)=0$ for $k\ne 0$. Therefore, we have two long-range dependent processes which are correlated but not cross-correlated. 

In a similar manner, let's have the same assumptions about correlations and parameters $\alpha, \beta, \gamma, \delta$ as in the previous two cases (only innovations $\varepsilon_2$ and $\varepsilon_3$ are correlated) but let's adjust the model specification to
\begin{gather}
x_t=\alpha\sum_{n=0}^{+\infty}{a_n(d_1)\varepsilon_{1,t-n}}+\beta\sum_{n=0}^{+\infty}\theta^n_2\varepsilon_{2,t-n} \nonumber \\
y_t=\gamma\sum_{n=0}^{+\infty}\theta^n_3\varepsilon_{3,t-n}+\delta\sum_{n=0}^{+\infty}{a_n(d_4)\varepsilon_{4,t-n}}.
\label{eq:ARFIMA_LC2}
\end{gather}
Processes $\{x_t\}$ and $\{y_t\}$ are thus linear combinations of ARFIMA(0,$d$,0) and AR(1) processes. In this case, we again have $\{x_t\}$ with memory $d_1$ (and $H_x=0.5+d_1$) and $\{y_t\}$ with memory $d_4$ (and $H_y=0.5+d_4$). And as the only non-zero correlation between innovations is $\sigma_{23}=\sigma_{32}$, we have $H_{xy}=0.5$ as the cross-correlations quickly vanish to zero. We thus have two long-range correlated processes $\{x_t\}$ and $\{y_t\}$, which are only short-range cross-correlated.

\begin{figure}[!htbp]
\begin{center}
\begin{tabular}{cc}
\includegraphics[width=70mm]{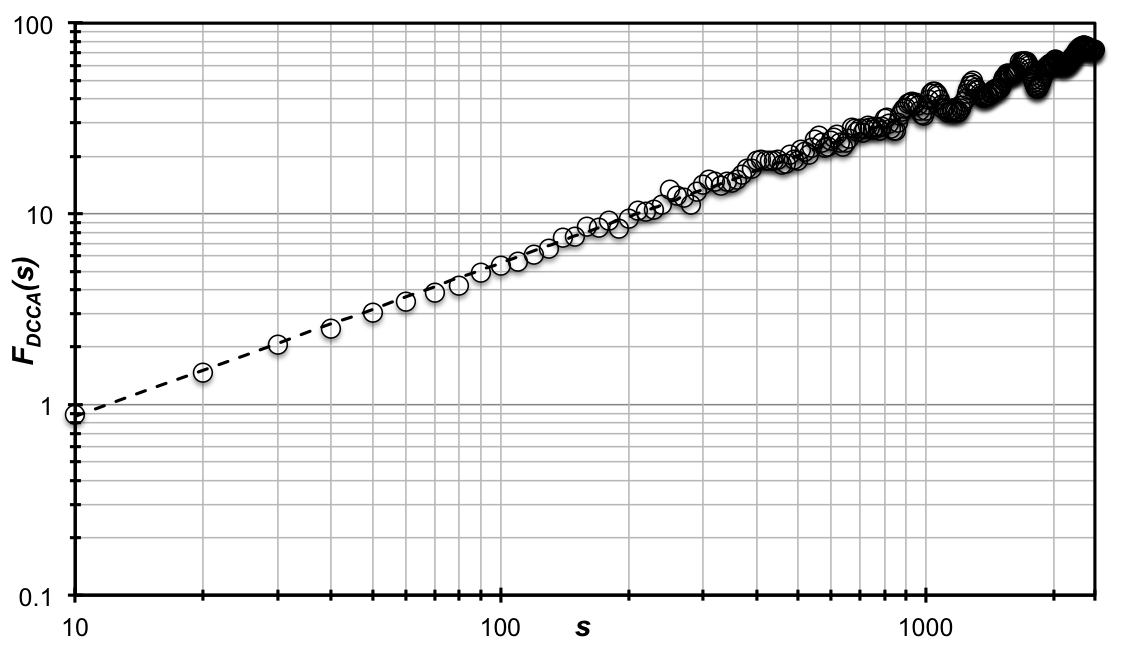}&\includegraphics[width=70mm]{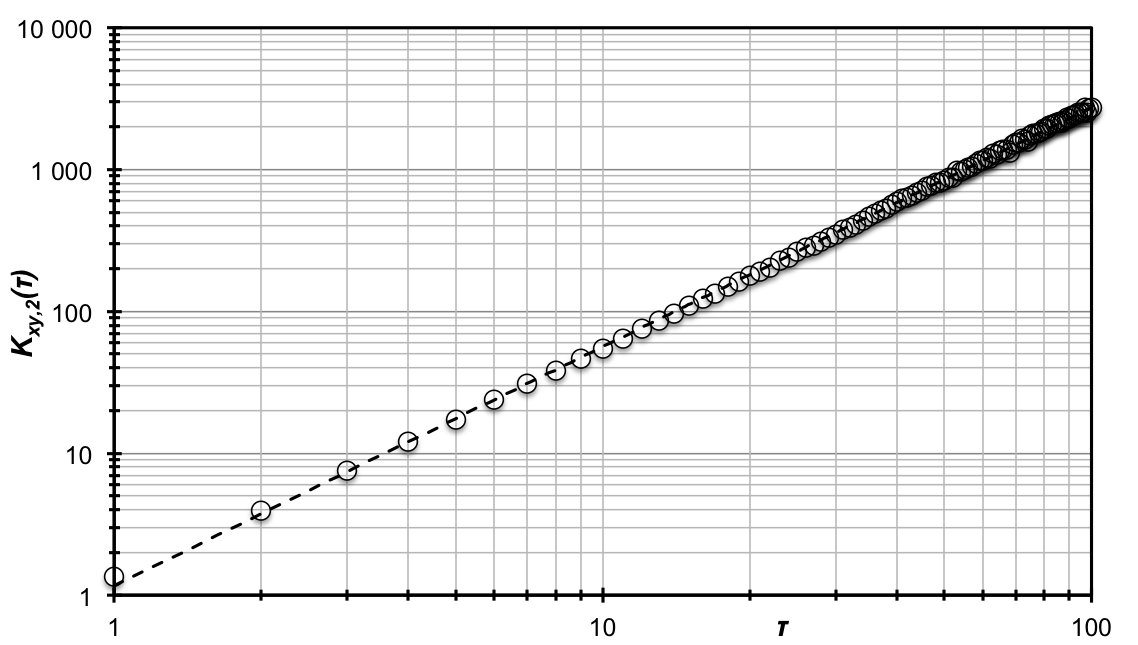}
\end{tabular}
\caption{\textbf{DCCA and HXA scaling for Model 1.} \footnotesize{The power-law scaling for both methods indicates long-range cross-correlations between series. The slope is in accordance with the theoretical $H_{xy}$.}\label{fig:Mixed_ARFIMA1_Scaling}}
\end{center}\vspace{-0.5cm}
\end{figure}

The MC-ARFIMA framework thus provides quite a wide range of possible model specifications yielding long-range cross-correlated processes which can be either long-range cross-correlated (with or without power law coherency), short-range cross-correlated, pairwise correlated or uncorrelated. Generally, the framework allows for even more possible specifications which encompass fractional cointegration and short-range cross-correlated AR processes to name the most interesting ones.

\section{Illustrative examples}

We present several simulated processes\footnote{R-project codes for MC-ARFIMA are available at \url{http://staff.utia.cas.cz/kristoufek/Ladislav_Kristoufek/Codes.html} or upon request from the author.} based on specifications developed in the previous section. For each of the processes, we show the cross-correlation function of a single simulation together with scatter plots of the series at selected lags (specifically lags 0, 1, 5 and 20) to see a different rate of decay of the selected processes.

\begin{figure}[!htbp]
\begin{center}
\begin{tabular}{cc}
\includegraphics[width=40mm]{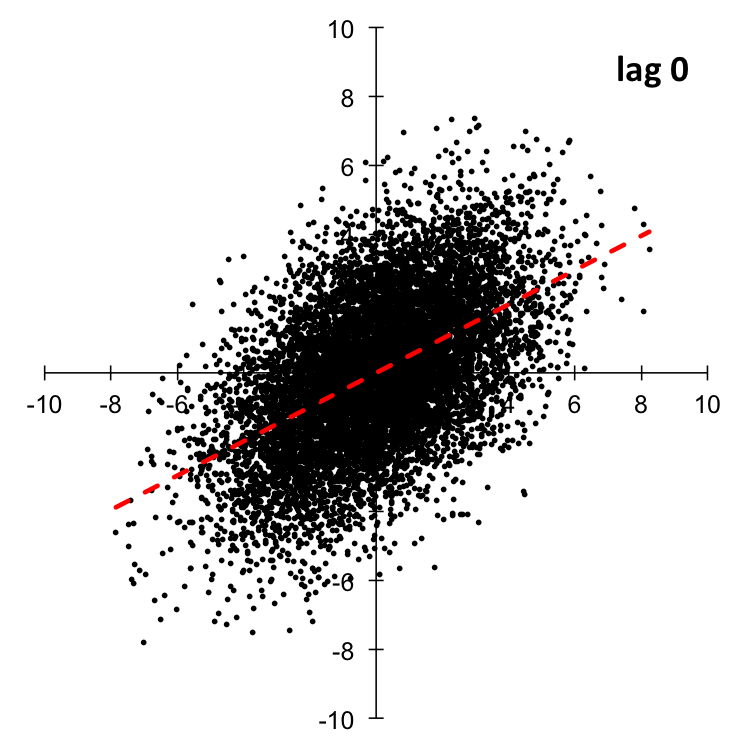}&\includegraphics[width=40mm]{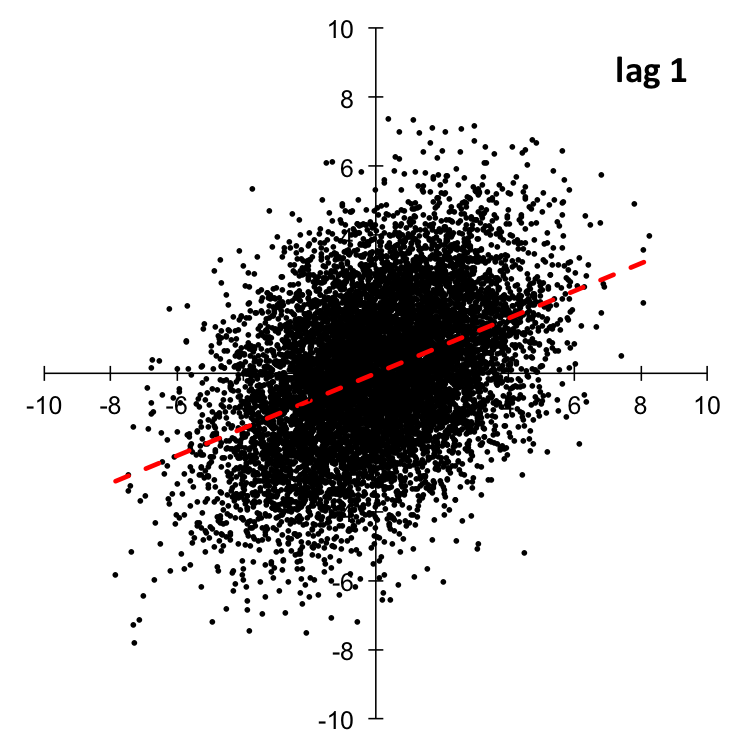}\\
\includegraphics[width=40mm]{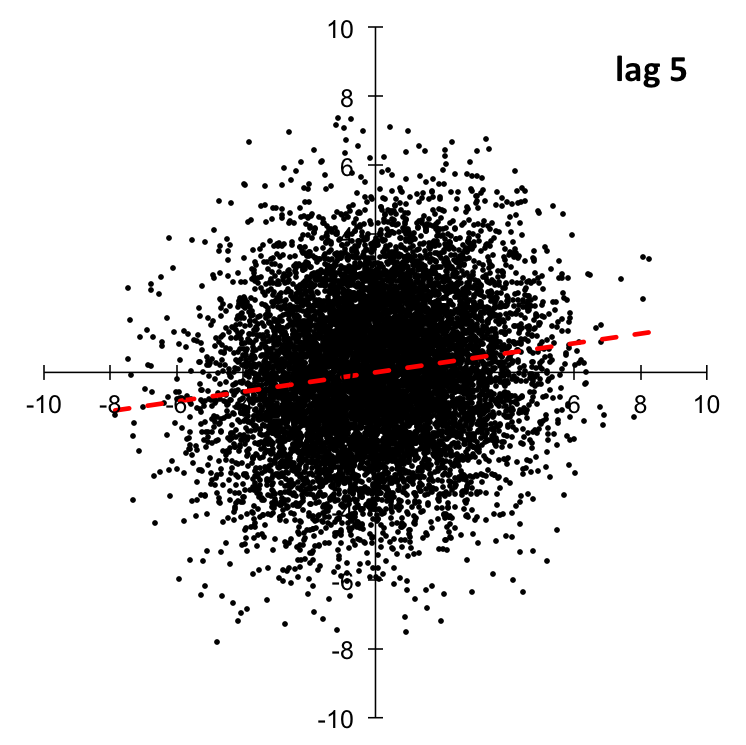}&\includegraphics[width=40mm]{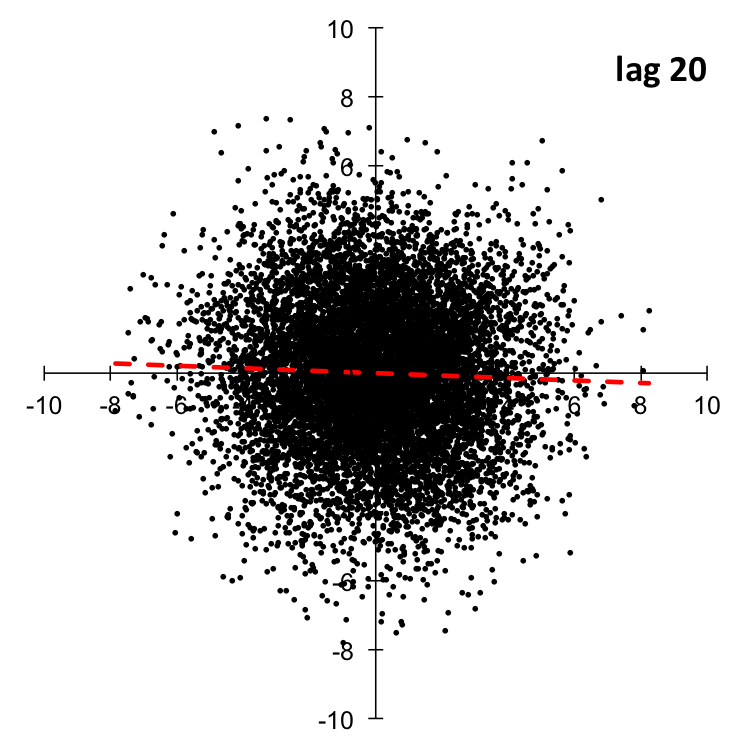}
\end{tabular}
\caption{\textbf{Cross-correlations decay of Model 2.} \footnotesize{Behavior of cross-correlations of selected lags illustrates a rapid decay. For lag five, the relationship between lagged series is still non-zero but for the twentieth lag, the relationship vanishes. Red dashed lines represent the least squares fits.} \label{fig:Mixed_ARFIMA2}}
\end{center}\vspace{-0.5cm}
\end{figure}

As Model 1, we use the specification of Eq. \ref{eq:ARFIMA_LC} with $\alpha=\delta=0.2$, $\beta=\gamma=1$, $d_1=d_4=0.4$, $d_2=d_3=0.3$, $\sigma^2_i=1$ for $i=1,2,3,4$ and $\sigma_{23}=0.9$ with $T=10000$. We thus obtain two long-range correlated processes with $H_x=H_y=0.9$ which are also long-range cross-correlated with $H_{xy}=0.8$. The values of parameters $\alpha,\beta,\gamma,\delta$ are selected to highlight the cross-persistence. The cross-correlation function between simulated processes is shown in Fig. \ref{fig:Mixed_ARFIMA1_CCF}. The cross-correlations evidently follow a very slow decay. The scatter plots (Fig. \ref{fig:Mixed_ARFIMA1}) for selected lags illustrate the decaying strength of the cross-correlations which, however, remain positive even for high lags. To see whether the decay can be identified as a power-law, we also present (Fig. \ref{fig:Mixed_ARFIMA1_Scaling}) scaling corresponding to the detrended cross-correlation analysis (DCCA)\footnote{DCCA is based on $s_{min}=10$, $s_{max}=T/5=2000$ with a step between scales of 10.} of Podobnik \textit{et al.} \citep{Podobnik2008} and the height cross-correlation analysis (HXA)\footnote{HXA is based on $\tau_{min}=1$ and $\tau_{max}=100$.} of Kristoufek \citep{Kristoufek2011}. Both methods show an apparent power-law scaling with a slope corresponding to $H_{xy}=0.8$. 

\begin{figure}[!htbp]
\begin{center}
\begin{tabular}{c}
\includegraphics[width=80mm]{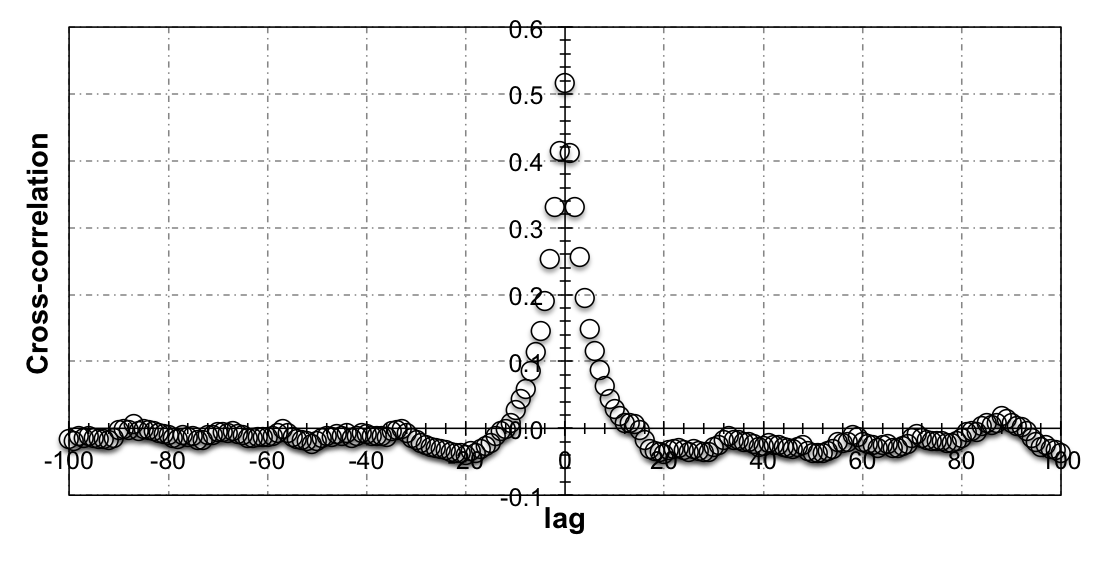}
\end{tabular}
\caption{\textbf{Cross-correlation function of Model 2.} \footnotesize{Cross-correlation function shows a rapid decay for both negative and positive lags and the cross-correlations vanish after approximately ten lags.} \label{fig:Mixed_ARFIMA2_CCF}}
\end{center}\vspace{-0.5cm}
\end{figure}

As Model 2, we use the specification given in Eq. \ref{eq:ARFIMA_LC2} with $\alpha=\beta=\gamma=\delta=1$, $d_1=d_4=0.4$, $\theta_2=\theta_3=0.8$, $\sigma^2_i=1$ for $i=1,2,3,4$ and $\sigma_{23}=0.9$ with $T=10000$. Model 2 thus represents two processes which are long-range correlated but only short-range cross-correlated. A rapid (exponential) decay of cross-correlation function is presented in Figs. \ref{fig:Mixed_ARFIMA2}-\ref{fig:Mixed_ARFIMA2_CCF}. Note that even for a rather strong short-term memory parameter $\theta=0.8$, the cross-correlations vanish quickly and after lag 15, the cross-correlations are close to zero. The exponential decay to insignificant cross-correlation values is more evident from the cross-correlation function.

\begin{figure}[!htbp]
\begin{center}
\begin{tabular}{cc}
\includegraphics[width=40mm]{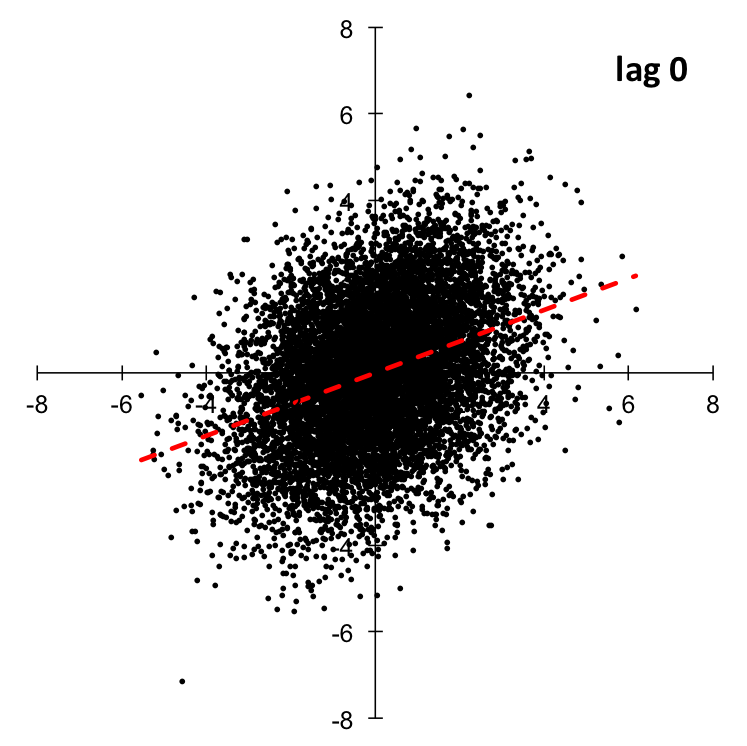}&\includegraphics[width=40mm]{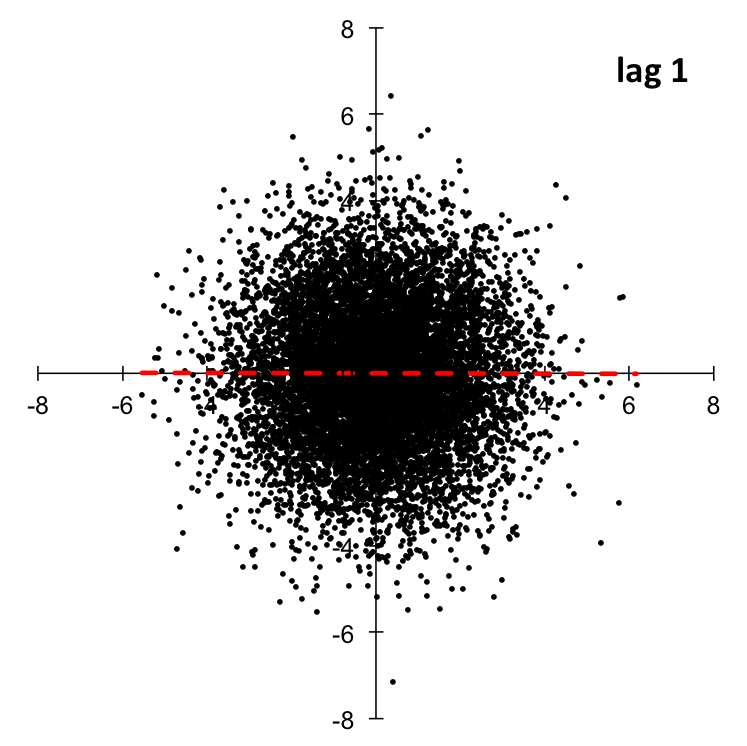}\\
\includegraphics[width=40mm]{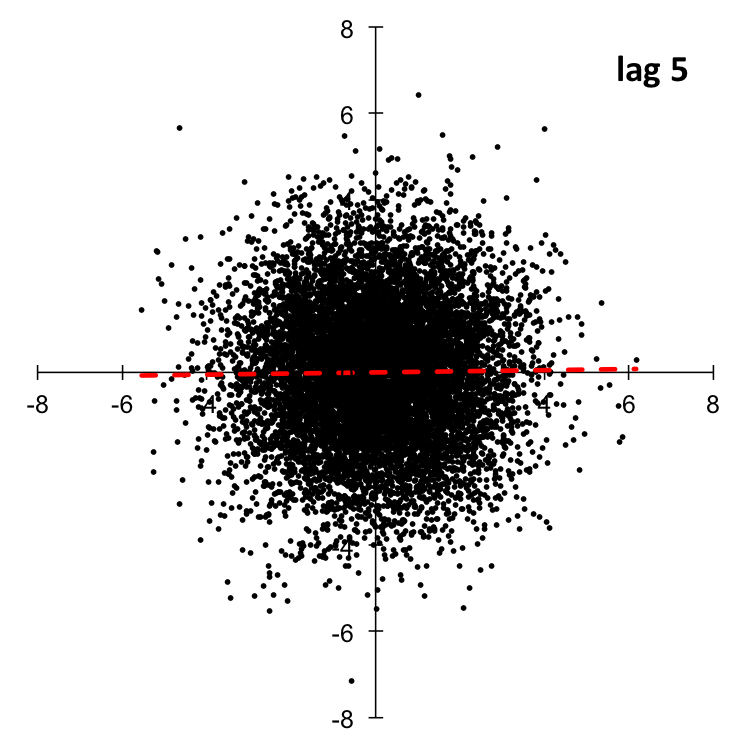}&\includegraphics[width=40mm]{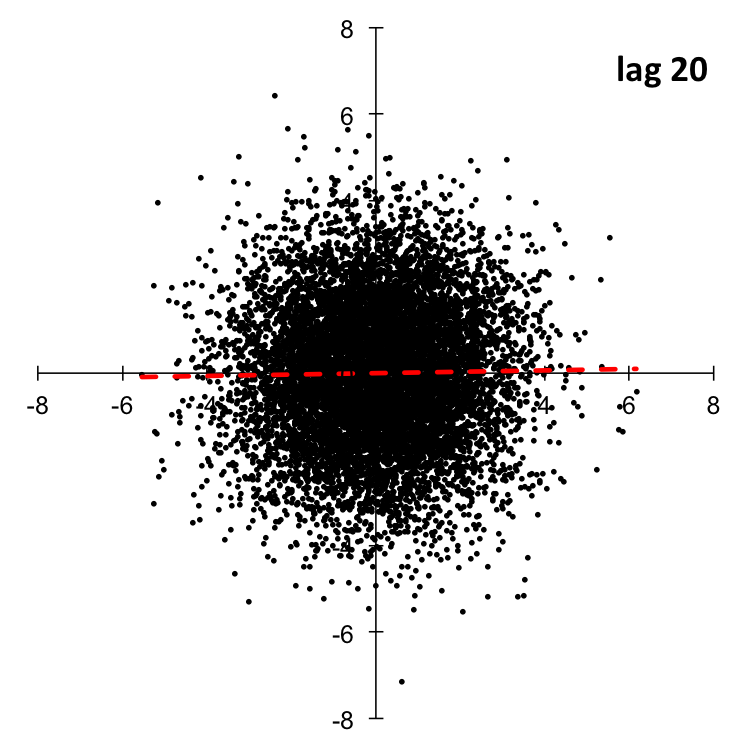}
\end{tabular}
\caption{\textbf{Cross-correlations decay of Model 3.} \footnotesize{Behavior of cross-correlations of selected lags illustrates the fact that the processes are only correlated but not cross-correlated. Apart from lag zero, there are no significant correlations between series. Red dashed lines represent the least squares fits.} \label{fig:Mixed_ARFIMA3}}
\end{center}\vspace{-0.5cm}
\end{figure}

\begin{figure}[!htbp]
\begin{center}
\begin{tabular}{c}
\includegraphics[width=80mm]{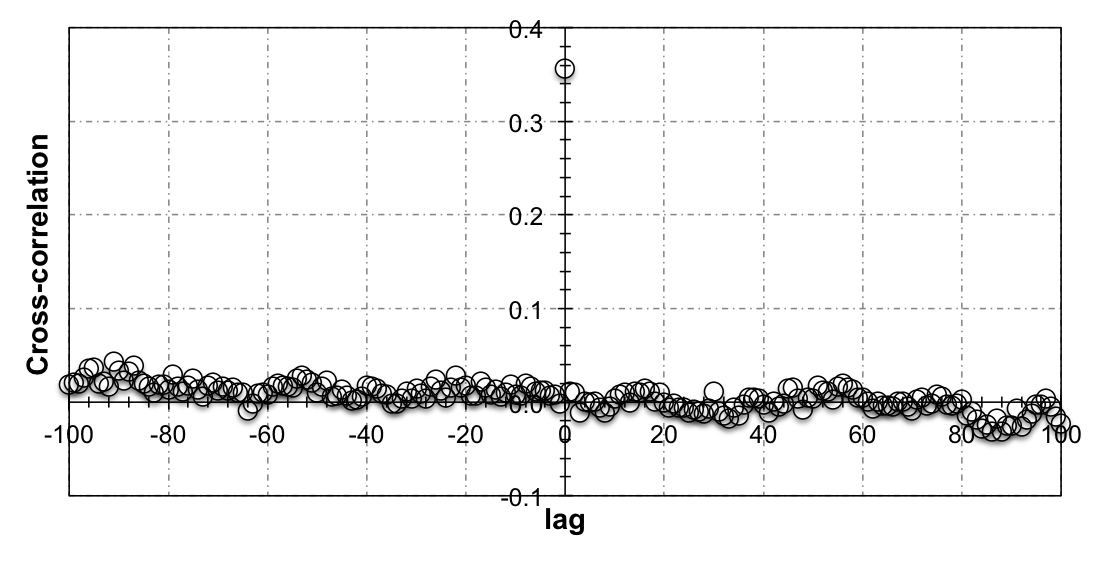}
\end{tabular}
\caption{\textbf{Cross-correlation function of Model 3.} \footnotesize{Cross-correlation function shows the only significant correlation at lag zero.} \label{fig:Mixed_ARFIMA3_CCF}}
\end{center}\vspace{-0.5cm}
\end{figure}

As Model 3, we use the specification given in Eq. \ref{eq:ARFIMA_Noise} with $\alpha=\beta=\gamma=\delta=1$, $d_1=d_4=0.4$, $\sigma^2_i=1$ for $i=1,2,3,4$ and $\sigma_{23}=0.9$ with $T=10000$. Model 3 thus represents two processes which are long-range correlated but only correlated (not cross-correlated). Form of the cross-correlation function is presented in Figs. \ref{fig:Mixed_ARFIMA3}-\ref{fig:Mixed_ARFIMA3_CCF}. We observe that indeed the cross-correlation function shows non-zero values only for the zeroth lag which is evident also from the scatter plots for specific lags.

\section{Discussion and conclusions}

Analysis of slowly decaying cross-correlations has recently become a widely and frequently discussed topic in the econophysics literature. However, only little attention has been put on the actual processes that can produce the bivariate Hurst exponent which is different from the average of the separate (univariate) Hurst exponents of the analyzed processes. In this paper, we have introduced a rather general framework of the Mixed-correlated ARFIMA (MC-ARFIMA) processes which allows for various specifications. Apart from a standard case when $H_{xy}=\frac{1}{2}(H_x+H_y)$, it also allows for processes with $H_{xy}<\frac{1}{2}(H_x+H_y)$ but also for long-range correlated processes which are either short-range cross-correlated or simply correlated. 

The major contribution of MC-ARFIMA lays in the fact that the processes have well-defined asymptotic properties for $H_x$, $H_y$ and $H_{xy}$ so that the processes can be used in simulation studies comparing various estimators of the bivariate Hurst exponent $H_{xy}$. Such a simulation study has not been conducted yet, leaving the question of statistical quality and accuracy (and performance in general) of the estimators unanswered. Importantly, the MC-ARFIMA framework allows for combination of short- and long-term memory properties so that various short-term memory biases can be studied. Indeed, the framework can be also expanded for different distributional properties of the innovations to study the performance of the estimators under heavy (possibly power-law) tails and asymmetry. 

Note, however, that the concept of having $H_{xy}<\frac{1}{2}(H_x+H_y)$ is new both in econophysics and financial econometrics so that processes with such property have not been studied adequately yet. In general, such property can be found between two series out of which at least one is power-law auto-correlated and the pair of series is power-law cross-correlated but the latter power-law decays faster than the former. Such behavior might be observed for example between interest rates and exchange rates of two countries. Interest rates of the two countries and their common exchange rate are expected to be tightly connected due to the interest rate parities (covered and uncovered) and such a relationship is expected to hold even for long scales. However, specifically in the times of financial turmoils (such as the current Global Financial Crisis), the cross-correlations tend to decay faster. Nonetheless, such a relationship should be examined in more detail with methods meant for testing or estimating the power-law cross-correlations. Other possible application comes in risk management where volatility is the variable of interest. It is well documented that volatility can be treated as a power-law auto-correlated process. Moreover, the volatility (or risk in general) has several components and it can be split between at least three specific risk factors -- market, industry and asset-specific ones. In the simplest form, the total volatility may be then seen as a linear combination of the three components and when a pair of financial assets is considered, the factors can be correlated across assets creating a possibility for the MC-ARFIMA modeling. As the financial assets (and economic/financial variables in general) are usually at least weakly correlated and many of them are power-law auto-correlated (fractionally integrated), possibilities for modeling are numerous and the two mentioned cases are set just as examples.

Last but not least, the framework allows for modeling of processes which are found to have $H_{xy}<\frac{1}{2}(H_x+H_y)$ which is novel in econophysics as well as in financial econometrics. Rather simple specification of the MC-ARFIMA processes enables to construct maximum likelihood estimators (MLE) as well as generalized method of moments (GMM) estimators and frequency-domain estimators.  

\section*{Acknowledgements}
The support from the Grant Agency of Charles University (GAUK) under project $1110213$, Grant Agency of the Czech Republic (GACR) under projects P402/11/0948 and 402/09/0965, and project SVV 267 504 are gratefully acknowledged.

\bibliography{Bibliography}
\bibliographystyle{chicago}

\section*{Appendix}
The pattern of the cross-correlation function for MC-ARFIMA defined in Eq. \ref{eq:ARFIMA_LC} is as follows:
\begin{multline}
\rho_{xy}(0)=\frac{\alpha\gamma\sigma_{13}}{\sigma_x\sigma_y}\sum_{k=0}^{+\infty}{a_k(d_1)a_k(d_3)}+\\
\frac{\alpha\delta\sigma_{14}}{\sigma_x\sigma_y}\sum_{k=0}^{+\infty}{a_k(d_1)a_k(d_4)}+\frac{\beta\gamma\sigma_{23}}{\sigma_x\sigma_y}\sum_{k=0}^{+\infty}{a_k(d_2)a_k(d_3)}+\frac{\beta\delta\sigma_{24}}{\sigma_x\sigma_y}\sum_{k=0}^{+\infty}{a_k(d_2)a_k(d_4)} \nonumber
\end{multline}
\begin{multline}
\rho_{xy}(1)=\frac{\alpha\gamma\sigma_{13}}{\sigma_x\sigma_y}\sum_{k=0}^{+\infty}{a_{k+1}(d_1)a_k(d_3)}+\\
\frac{\alpha\delta\sigma_{14}}{\sigma_x\sigma_y}\sum_{k=0}^{+\infty}{a_{k+1}(d_1)a_k(d_4)}+\frac{\beta\gamma\sigma_{23}}{\sigma_x\sigma_y}\sum_{k=0}^{+\infty}{a_{k+1}(d_2)a_k(d_3)}+\frac{\beta\delta\sigma_{24}}{\sigma_x\sigma_y}\sum_{k=0}^{+\infty}{a_{k+1}(d_2)a_k(d_4)} \nonumber
\end{multline}
\begin{gather*}
\vdots \nonumber
\end{gather*}
\begin{multline}
\rho_{xy}(i)=\frac{\alpha\gamma\sigma_{13}}{\sigma_x\sigma_y}\sum_{k=0}^{+\infty}{a_{k+i}(d_1)a_k(d_3)}+\\
\frac{\alpha\delta\sigma_{14}}{\sigma_x\sigma_y}\sum_{k=0}^{+\infty}{a_{k+i}(d_1)a_k(d_4)}+\frac{\beta\gamma\sigma_{23}}{\sigma_x\sigma_y}\sum_{k=0}^{+\infty}{a_{k+i}(d_2)a_k(d_3)}+\frac{\beta\delta\sigma_{24}}{\sigma_x\sigma_y}\sum_{k=0}^{+\infty}{a_{k+i}(d_2)a_k(d_4)} \nonumber
\end{multline}
\begin{multline}
\rho_{xy}(-1)=\frac{\alpha\gamma\sigma_{13}}{\sigma_x\sigma_y}\sum_{k=0}^{+\infty}{a_{k}(d_1)a_{k+1}(d_3)}+\\
\frac{\alpha\delta\sigma_{14}}{\sigma_x\sigma_y}\sum_{k=0}^{+\infty}{a_{k}(d_1)a_{k+1}(d_4)}+\frac{\beta\gamma\sigma_{23}}{\sigma_x\sigma_y}\sum_{k=0}^{+\infty}{a_{k}(d_2)a_{k+1}(d_3)}+\frac{\beta\delta\sigma_{24}}{\sigma_x\sigma_y}\sum_{k=0}^{+\infty}{a_{k}(d_2)a_{k+1}(d_4)} \nonumber
\end{multline}
\begin{gather}
\vdots
\label{eq:MC-ARFIMA_rho}
\end{gather}
\begin{multline}
\rho_{xy}(-i)=\frac{\alpha\gamma\sigma_{13}}{\sigma_x\sigma_y}\sum_{k=0}^{+\infty}{a_{k}(d_1)a_{k+i}(d_3)}+\\
\frac{\alpha\delta\sigma_{14}}{\sigma_x\sigma_y}\sum_{k=0}^{+\infty}{a_{k}(d_1)a_{k+i}(d_4)}+\frac{\beta\gamma\sigma_{23}}{\sigma_x\sigma_y}\sum_{k=0}^{+\infty}{a_{k}(d_2)a_{k+i}(d_3)}+\frac{\beta\delta\sigma_{24}}{\sigma_x\sigma_y}\sum_{k=0}^{+\infty}{a_{k}(d_2)a_{k+i}(d_4)} \nonumber
\end{multline}
Evidently, the cross-correlations structure does not depend on time $t$ and together with the wide-sense stationarity of separate processes $\{x_t\}$ and $\{y_t\}$, MC-ARFIMA processes are also jointly wide-sense stationary.
\end{document}